\documentclass[12pt]{article}
\usepackage{amsmath}
\usepackage{graphicx}
\usepackage{enumerate}
\usepackage{natbib}
\usepackage{url} % not crucial - just used below for the URL
\usepackage{booktabs}

\newcommand{\blind}{1}

\begin{document}

\def\spacingset#1{\renewcommand{\baselinestretch}%
{#1}\small\normalsize} \spacingset{1}

%%%%%%%%%%%%%%%%%%%%%%%%%%%%%%%%%%%%%%%%%%%%%%%%%%%%%%%%%%%%%%%%%%%%%%%%%%%%%%

\if1\blind
{
  \title{\bf A novel Phase I clinical trial design with unequal cohort sizes}
  \author{Xiaojun Zhu\thanks{
    The authors gratefully acknowledge \textit{Natural Science Foundation of Jiangsu Province, China-The Excellent Young Scholar Programme [No. BK20220098] and Research Enhancement Fund of Xi'an-Jiaotong Liverpool University [No. REF-22-01-012]}}\hspace{.2cm}\\
    School of Mathematics and Physics, Xi'an Jiaotong-Liverpool University}
  \maketitle
} \fi

\if0\blind
{
  \bigskip
  \bigskip
  \bigskip
  \begin{center}
    {\LARGE\bf A novel Phase I clinical trial design with unequal cohort sizes}
\end{center}
  \medskip
} \fi

\bigskip
\begin{abstract}
This paper introduces a new Phase I design aimed at enhancing the performance of existing methods, including algorithm-based, model-based, and model-assisted designs.
The design, developed by integrating the concept of Fisher information, is easily operationalized.
The new design addresses the issue of the classical designs' slow dosage escalation. 
Simulation demonstrate that the proposed design markedly enhances performance in terms of efficiency, accuracy, and reliability. 
Moreover, the trial duration has been notably reduced with a large sample size.
\end{abstract}

\noindent%
\vfill

\newpage
\spacingset{1.9} % DON'T change the spacing!
\section{Introduction}
\label{sec:intro}
In Phase I trial, one of the primary objectives is to determine the maximum tolerated dose (MTD) of a new drug,
as well as to evaluate its safety, pharmacokinetics, and physiological effects.
The MTD is the dose most likely to induce specified levels of toxicity.
Dose-limiting toxicities (DLTs) are drug-induced toxicities that prevent further treatment.
The trial typically involves a small sample size of 20 to 50 subjects, usually treated in cohorts of 3, and begins with the lowest or doctor-recommended dosage.
Dose adjustments are made in a sequential and adaptive manner based on observed DLTs: with lower doses administered to the next cohort if too many DLTs occur, 
and higher dosages administered if the current dose is well-tolerated.

Many statistical methods have been developed to find the MTD, generally classified as algorithm-based, model-based, and model-assisted designs.
The algorithm-based designs include 3+3 and its generalization i3+3 \cite{Liu3}, rolling six designs \cite{Sto,Sko}, the biased-coin design \cite{Dur} and its extensions \cite{Sty, Iva},  and accelerated titration design \cite{Sim}.
Due to the transparency and simplicity, algorithm-based designs are widely used in practice \cite{Rog}.
However, the performance is usually not very well because these designs are model-free, i.e., non-parametric based.
Model-based designs include the continual reassessment method (CRM) \cite{O'qu} and its various extensions
such as
dose escalation with overdose control \cite{Bab}, time-to-event CRM on late-onset toxicities \cite{Che},
Bayesian model averaging CRM \cite{Yin},
Bayesian data-augmentation CRM \cite{Liu},
partial order CRM \cite{Wag},
quasi-partial order CRM \cite{O'co},
bivariate CRM on two competing outcomes \cite{Bra},
pediatric-CRM \cite{Li} etc.
Interested readers may refer to \cite{Che2} for a detailed reviews on CRM.
Recently, model-assisted designs have become popular, such as the
modified toxicity probability interval \cite{Ji} and the related extensions \cite{Guo},
Bayesian optimal interval design \cite{Liu2, Yua} and its extensions \cite{Lin, Yua2, Koj},
Keyboard designs \cite{Yan} and its variation on delayed toxicity \cite{Lin2}.
One may refer to \cite{Yin2, Hor, Pet, Zho, Che3} on the  reviews and comparisons of different Phase I design methods.

There is limited discourse in the literature regarding the cohort size in Phase I design. 
``The number of patients to be included in each cohort`` is one of the top 10 questions that should be deliberated upon when designing dose-finding Phase I clinical trials; refer to \cite{Lee2021}.
To fill this gap, the aim of this manuscript is to propose a novel design with unequal cohort sizes to enhance trial performance. 
The design with unequal cohort sizes under consideration can be readily incorporated into most existing designs.

The rest of the paper is organized as follows.
Section 2 provides a concise overview of the statistical rationale employed to ascertain the cohort size across various time periods. 
Section 3 assesses the performance of implementing unequal cohort sizes in the context of CRM and Keyboard through a simulation study. 
Section 4 encompasses discussions and conclusions.

\section{Unequal Cohort Sizes}
Intuitively, the information gained from early cohorts is limited, so the cohort size should be small due to uncertainty.
This approach also benefits the patients undergoing treatment in the early stages.
As information accumulates and uncertainty reduces, the confidence to increase the cohort size grows in turn.
This motivates us to design the cohort sizes based on Fisher information.

Let $d_1, d_2, \cdots, d_J$ denote $J$ preplanned doses arranged in ascending order, with toxicity rates denoted as $p_1$ to $p_J$.
Now, assume $n_j$ patients have been treated in dose $j$, then the Fisher information is
$$
\sum\limits_{j=1}^J \frac{n_j}{p_j(1-p_j)}.
$$
The information obtained at each dose level $j$ is proportional to $n_j$ and $1/[p_j(1-p_j)]$.

After several cohorts, the recommended dosage for the next cohort stabilizes and is consistently the one with a toxicity rate near the target level, say $\phi$.
Sometimes, the recommended dosage may not be the optimal one, but it still has a toxicity rate close to the desired level; see \cite{She, Che4}.
Define $N=\sum\limits_{i=1}^J n_j$, then the information will be approach to 
$$
\sum\limits_{j=1}^J \frac{n_j}{\phi(1-\phi)}=\frac{1}{\phi(1-\phi)} \sum\limits_{j=1}^J n_j=\frac{N}{\phi(1-\phi)}.
$$
Here,  the closeness is based on how toxic the MTD or doses near the MTD are in a few cases.
As the information is roughly proportional to $N$, motivating us to use a cohort sizes that are related to $N^{-1/2}$, i.e.,
from the  perspective of  standard deviation.

For illustration, let us assume we have a trial with 30 patients.
Table \ref{table1} presents values of $N^{1/2}$ and $1/[N^{1/2}]$ for $N$ from $1$ to $30$, where $[\cdot]$ denotes the rounding function.
According to the values of $1/[N^{1/2}]$, we propose to use the cohort sizes $1,1,2,2,3,3,4$,
$4,5,5$, making a total of 10 cohorts, which is the same as if we use fixed cohort sizes kept as 3.  
The simulation study carried out in the next Section,
demonstrates that using unequal cohort sizes performs better than using equal cohort sizes.
\begin{table}
\caption{Values of $N^{1/2}$ and $1/[N^{1/2}]$ for $N$ from $1$ to $30$.}
\label{table1}
\centering
\begin{tabular}{ccccccccccc}
\toprule
$N$ & 1 & 2 & 3 & 4 & 5 & 6 & 7 & 8 & 9 & 10\\
$N^{1/2}$ & 1.00 & 1.41 & 1.73 & 2.00 & 2.24 & 2.45 & 2.65 & 2.83 & 3.00 & 3.16\\
$1/[N^{1/2}]$ & 1/1 & 1/1 & 1/2 & 1/2 & 1/2 & 1/2 & 1/3 & 1/3 & 1/3 & 1/3 \\
$N$ & 11 & 12 & 13 & 14 & 15 & 16 & 17 & 18 & 19 & 20\\
$N^{1/2}$  & 3.32 & 3.46 & 3.61 & 3.74 & 3.87 & 4.00 & 4.12 & 4.24 & 4.36 & 4.47\\
$1/[N^{1/2}]$ & 1/3 & 1/3 & 1/4 & 1/4 & 1/4 & 1/4 & 1/4 & 1/4 & 1/4 & 1/4 \\
 $N$ & 21 & 22 & 23 & 24 & 25 & 26 & 27 & 28 & 29 & 30\\
$N^{1/2}$  & 4.58 & 4.69 & 4.80 & 4.90 & 5.00 & 5.10 & 5.20 & 5.29 & 5.39 & 5.48\\
$1/[N^{1/2}]$ & 1/5 & 1/5 & 1/5 & 1/5 & 1/5 & 1/5 & 1/5 & 1/5 & 1/5 & 1/5 \\
\bottomrule
\end{tabular}
\end{table}

In general, we recommend to use the cohort sizes $1,1,2,2,3,3,4,4,5,5,6,6,\cdots$, i.e., the size in the $i$-th cohort equals to $[i/2]$. 
In practice, one could slightly adjust the last cohort size according to the total sample size $N$.
For example, if the total size is $N=24$, one can have the cohort sizes $1,1,2,2,3,3,4,4,4$;
if the total size is $N=26$, one  can have the cohort sizes $1,1,2,2,3,3,4,4,6$.

\section{Simulation Study}
To evaluate the performance of the proposed cohort sizes and fixed cohort sizes, we carry out a Monte Carlo simulation study.
For the sake of brevity, only comparisons on CRM and Keyboard designs are presented.
One could also incorporate the proposed cohort sizes into other designs.
The total sample sizes are chosen as $N=24$, $30$, $36$ and $42$.
The target toxic probabilities are $\phi=0.3$ for all scenarios.
For CRM, the power model $\pi_j(\alpha)=p_j^{\exp(\alpha)}$, where $\alpha\sim n(0,1.34)$ and
the skeleton $(0.1, 0.2, 0.3, 0.4, 0.5, 0.6)$ are used.
For Keyboard,  the appropriate dosing interval is chosen as (25$\%$, 35$\%$).
For safety, the trial starts from the lowest dose level and 	
requires no dose skipping during dose escalation or de-escalation.

The cohort-related information is summarized in Table \ref{table2}.
If the total number of patients $N$ exceeds 30, having unequal cohort sizes can shorten the trial.
If there are more than 132 patients, using unequal cohort sizes reduce the number of cohorts required by at least half.

\begin{table}
\centering
\caption{Cohort-related information.}
\label{table2}
\begin{tabular}{lcc}
\toprule
 & Unequal Cohort Sizes & Fixed Cohort Sizes \\
\midrule
\multicolumn{3}{c}{24 Patients}\\
Cohort Sizes & 1,1,2,2,3,3,4,4,4 & 3,3,3,3,3,3,3,3\\
Total Cohorts &  9 & 8\\
\midrule
\multicolumn{3}{c}{30 Patients}\\
Cohort Sizes & 1,1,2,2,3,3,4,4,5,5 & 3,3,3,3,3,3,3,3,3,3\\
Total Cohorts &  10 & 10\\
\midrule
\multicolumn{3}{c}{36 Patients}\\
Cohort Sizes & 1,1,2,2,3,3,4,4,5,5,6 & 3,3,3,3,3,3,3,3,3,3,3,3\\
Total Cohorts &  11 & 12\\
\midrule
\multicolumn{3}{c}{42 Patients}\\
Cohort Sizes & 1,1,2,2,3,3,4,4,5,5,6,6 & 3,3,3,3,3,3,3,3,3,3,3,3,3,3\\
Total Cohorts &  12 & 14\\
\bottomrule
\end{tabular}
\end{table}

Using unequal cohort sizes (denoted as the new method)  in CRM and Keyboard generally performs better in accurately selecting the MTD compared to using fixed cohort sizes.

The performance of CRM, Keyboard  with unequal cohort sizes (denoted as a new method) and fixed cohort sizes are reported in Tables \ref{table3}-\ref{table10}, respectively.
For CRM,  the design with unequal cohort sizes shows higher accuracy  (MTD selection $\%$) in all scenarios 
and allocates more patients to the MTD in scenarios 2-6 when $N=24$.
When $N=30, 36$, $42$, the new design performs almost identically as the classical one when the MTD is between the lowest two dosage,
but usually has a better performance in other cases, especially when the MTD is the largest dosage.
Moreover,  the new design always has a lower overdose rate recommendation except for the case when MTD is  the lowest dosage.
However, the new design is more easily to allocate 1-2 patients on a dose higher the MTD.
For Keyboard, the new design usually has a good performance in accuracy unless the MTD is the lowest or the largest dosage.
For Scenario 6, the classical design has a better performance in accuracy when $N=30$, $36$ and $42$.
This is because the chance of toxicity for dose 5 is 0.23, which is close to the target level, so the new design suggests this dosage as the MTD. 
However, it should be noted that the new design results in many more patients receiving doses 5 and 6 for this scenario. 
Regarding the overdose rate recommendation, the new design outperforms the classical one for scenarios 3-5, and both designs have a similar 
performance for scenario 2.
Similar to the CRM case, the new design will more easily to recommend a higher dosage as the MTD if the lowest dose is actually the MTD.
Once again, the Keyboard with unequal cohort sizes is more likely to allocate 1-2 more patients to a dose higher than the MTD.

\spacingset{1} 
\begin{table}
\centering
\caption{Performance of CRM with unequal and fixed cohort sizes when total size is $N=24$. }
\label{table3}
\begin{tabular}{llrrrrrrr}
\hline
    &  & \multicolumn{6}{c}{Dose Level} &  \multicolumn{1}{c}{Overdose}\\
    & & & & & & & &    \multicolumn{1}{c}{Rate} \\
    \midrule
    &  & \multicolumn{1}{c}{1} & \multicolumn{1}{c}{2}
    & \multicolumn{1}{c}{3} & \multicolumn{1}{c}{4}
    & \multicolumn{1}{c}{5} & \multicolumn{1}{c}{6}\\
Scenario 1 &  &  {\bf 0.30} & 0.38 & 0.48 & 0.58 & 0.69 & 0.78\\
CRM & Selection (\%) & 65.71 & 26.68 & 6.80 & 0.76 & 0.05 & 0.00 & 34.29\\
    & Patient  (\%) & 15.97 &  5.32 &  2.27 &  0.40 &  0.02 & 0.00 & 8.03\\
New CRM & Selection (\%) & 65.84 & 26.58 & 7.01 & 0.56 & 0.01 & 0.00 & 34.16\\
    & Patient  (\%) & 13.79 & 5.74 &  3.27 &  1.04 &  0.14  & 0.02 & 10.21\\
    \\
Scenario 2 &  & 0.20 & {\bf 0.30} & 0.45 & 0.55 & 0.60 & 0.70\\
CRM & Selection (\%) & 29.74 & 47.68 & 19.37 & 2.86 & 0.34 & 0.01 & 22.58\\
    & Patient  (\%) & 10.37 & 7.91 &  4.52 &  1.05 & 0.14 & 0.01 & 5.72\\
New CRM & Selection (\%) & 30.95 & 49.48 & 17.47 & 1.88 & 0.21 & 0.01 & 19.57\\
    & Patient  (\%) & 8.67 & 8.13 & 5.17 & 1.71 & 0.27 & 0.05 & 7.20\\
    \\
Scenario 3 &  & 0.05 & 0.10 & {\bf 0.30} & 0.50 & 0.65 & 0.75\\
CRM & Selection (\%) & 0.06 & 14.26 & 61.42 & 22.50 & 1.74 & 0.02 & 24.26\\
    & Patient  (\%) & 3.70 & 5.15&  9.24 &  5.01 & 0.85 & 0.05 & 5.90\\
New CRM & Selection (\%) & 0.26 & 18.05 & 65.06 & 15.87 & 0.76 & 0.00 & 16.63\\
    & Patient  (\%) & 1.82 &  5.25 & 10.64 & 5.29 &  0.86 &  0.13 & 6.29\\
        \\
Scenario 4 &  & 0.07 & 0.12 & 0.17 & {\bf 0.30} & 0.45 & 0.60\\
CRM & Selection (\%) & 0.28 & 4.50 & 24.67 & 47.01 & 21.12 & 2.42 & 23.54\\
    & Patient  (\%) & 4.05 & 4.08 & 6.04 & 6.39 & 2.89 & 0.56 & 3.45\\
New CRM & Selection (\%) & 0.27 & 4.40 & 28.46 & 47.22 & 18.29 & 1.36 & 19.65\\
    & Patient  (\%) & 2.03 & 2.76 & 6.30 & 7.93 & 3.83 & 1.15 & 4.98\\
        \\
Scenario 5 &  & 0.04 & 0.08 & 0.12 & 0.15 & {\bf 0.30} & 0.50\\
CRM & Selection (\%) & 0.01 & 0.65 & 5.15 & 23.03 & 51.21 & 19.95 & 19.95\\
    & Patient  (\%)  & 3.48 & 3.34 & 4.19 & 5.13 & 5.46 & 2.40 & 2.40\\
New CRM & Selection (\%) & 0.00 & 0.55 & 5.20 & 26.30 & 53.12 & 14.83 & 14.83\\
        & Patient  (\%) & 1.51 & 1.76 & 3.64 & 6.03 & 7.07 & 3.98 & 3.98\\
        \\
Scenario 6 &  & 0.05 & 0.14 & 0.18 & 0.20 & 0.23 & {\bf 0.30}\\
CRM & Selection (\%) & 0.14 & 4.37 & 14.78 & 23.43 & 29.88 & 27.40 & 0.00\\
    & Patient  (\%)  & 3.77 & 4.09 & 5.44 & 4.87 & 3.62 & 2.21 & 0.00\\
New CRM & Selection (\%) & 0.13 & 3.79 & 12.84 & 19.87 & 28.75 & 34.62 & 0.00\\
        & Patient  (\%) & 1.85 & 2.71 & 4.74 & 5.30 & 4.50 & 4.91 & 0.00\\
\hline
\end{tabular}
\end{table}

\begin{table}
\centering
\caption{Performance of CRM with unequal and fixed cohort sizes when total size is $N=30$. }
\label{table4}
\begin{tabular}{llrrrrrrr}
\hline
    &  & \multicolumn{6}{c}{Dose Level} &  \multicolumn{1}{c}{Overdose}\\
    & & & & & & & &    \multicolumn{1}{c}{Rate} \\
      \midrule
    &  & \multicolumn{1}{c}{1} & \multicolumn{1}{c}{2}
    & \multicolumn{1}{c}{3} & \multicolumn{1}{c}{4}
    & \multicolumn{1}{c}{5} & \multicolumn{1}{c}{6}\\
Scenario 1 &  &  {\bf 0.30} & 0.38 & 0.48 & 0.58 & 0.69 & 0.78\\
CRM & Selection (\%) & 66.14 & 28.25 & 5.17 & 0.44 & 0.00 & 0.00 & 33.86\\
    & Patient  (\%) & 19.83 &  6.94 &  2.70 &  0.47 &  0.05 &  0.00 & 10.17\\
New CRM & Selection (\%) & 65.78 & 28.40 & 5.42 & 0.40 & 0.00 & 0.00 & 34.22\\
    & Patient  (\%) & 17.54 & 7.47 & 3.72 & 1.12 & 0.13 & 0.02 & 12.46\\
    \\
Scenario 2 &  & 0.20 & {\bf 0.30} & 0.45 & 0.55 & 0.60 & 0.70\\
CRM & Selection (\%) & 26.74 & 54.77 & 17.12 & 1.32 & 0.05 & 0.00 & 18.49\\
    & Patient  (\%) & 12.05 & 11.09 & 5.59 & 1.13 &  0.14 & 0.01 & 6.86\\
New CRM & Selection (\%) & 27.73 & 53.63 & 17.19 & 1.41 & 0.04 & 0.00 & 18.64\\
    & Patient  (\%) & 10.34 & 11.14 & 6.33 & 1.84 & 0.29 & 0.06 & 8.52\\
    \\
Scenario 3 &  & 0.05 & 0.10 & {\bf 0.30} & 0.50 & 0.65 & 0.75\\
CRM & Selection (\%) & 0.01 & 13.07 & 67.16 & 19.22 & 0.54 & 0.00 & 19.76\\
    & Patient  (\%) & 3.69 & 6.02 & 13.08 & 6.21 & 0.94 & 0.05 & 7.21\\
New CRM & Selection (\%) & 0.07 & 15.22 & 68.64 & 15.67 & 0.40 & 0.00 & 16.07\\
    & Patient  (\%) & 1.83 & 6.43 & 14.37 & 6.36 & 0.88 & 0.14 & 7.38\\
        \\
Scenario 4 &  & 0.07 & 0.12 & 0.17 & {\bf 0.30} & 0.45 & 0.60\\
CRM & Selection (\%) & 0.07 & 2.98 & 23.26 & 53.23 & 19.31 & 1.15 & 20.46\\
    & Patient  (\%)  & 4.08 & 4.28 & 7.45 & 9.26 & 4.24 & 0.69 & 4.93\\
New CRM & Selection (\%) & 0.07 & 2.55 & 23.85 & 53.02 & 19.60 & 0.91 & 20.51\\
    & Patient  (\%) & 1.99 & 2.93 & 7.77 & 10.99 & 5.07 & 1.27 & 6.33\\
        \\
Scenario 5 &  & 0.04 & 0.08 & 0.12 & 0.15 & {\bf 0.30} & 0.50\\
CRM & Selection (\%) & 0.00 & 0.33 & 3.41 & 23.59 & 56.95 & 15.72 & 15.72\\
    & Patient  (\%)  & 3.50 & 3.41 & 4.52 & 6.63 & 8.53 & 3.43 & 3.43\\
New CRM & Selection (\%) & 0.00 & 0.20 & 3.38 & 22.81 & 59.68 & 13.93 & 13.93\\
        & Patient  (\%) & 1.51 & 1.80 & 3.93 & 7.52 & 10.24 & 5.00 & 5.00\\
        \\
Scenario 6 &  & 0.05 & 0.14 & 0.18 & 0.20 & 0.23 & {\bf 0.30}\\
CRM & Selection (\%) & 0.06 & 2.86 & 13.00 & 24.14 & 29.71 & 30.23 & 0.00\\
    & Patient  (\%)  & 3.81 & 4.35 & 6.34 & 6.24 & 5.35 & 3.91 & 0.00\\
New CRM & Selection (\%) & 0.06 & 3.04 & 10.91 & 19.39 & 29.05 & 37.55 & 0.00\\
        & Patient  (\%) & 1.93 & 3.07 & 5.58 & 6.34 & 6.03 & 7.05 & 0.00\\
\hline
\end{tabular}
\end{table}

\begin{table}
\centering
\caption{Performance of CRM with unequal and fixed cohort sizes when total size is $N=36$. }
\label{table5}
\begin{tabular}{llrrrrrrr}
\hline
    &  & \multicolumn{6}{c}{Dose Level} &  \multicolumn{1}{c}{Overdose}\\
    & & & & & & & &    \multicolumn{1}{c}{Rate} \\
      \midrule
    &  & \multicolumn{1}{c}{1} & \multicolumn{1}{c}{2}
    & \multicolumn{1}{c}{3} & \multicolumn{1}{c}{4}
    & \multicolumn{1}{c}{5} & \multicolumn{1}{c}{6}\\
Scenario 1 &  &  {\bf 0.30} & 0.38 & 0.48 & 0.58 & 0.69 & 0.78\\
CRM & Selection (\%) & 69.59 & 26.26 & 3.99 & 0.16 & 0.00 & 0.00 & 30.41\\
    & Patient  (\%) & 24.05 & 8.55 & 2.91 & 0.45 & 0.04 & 0.00 & 11.95\\
New CRM & Selection (\%) & 67.08 & 28.17 & 4.58 & 0.17 & 0.00 & 0.00 & 32.92\\
    & Patient  (\%) & 21.61 & 9.10 & 4.08 & 1.06 & 0.13 & 0.02 & 14.39\\
    \\
Scenario 2 &  & 0.20 & {\bf 0.30} & 0.45 & 0.55 & 0.60 & 0.70\\
CRM & Selection (\%) & 25.52 & 57.57 & 16.06 & 0.82 & 0.03 & 0.00 & 16.91\\
    & Patient  (\%) & 13.75 & 14.31 & 6.57 & 1.23 & 0.13 & 0.01 & 7.94\\
New CRM & Selection (\%) & 25.89 & 56.73 & 16.42 & 0.94 & 0.02 & 0.00 & 17.38\\
    & Patient  (\%) & 12.26 & 14.18 & 7.27 & 1.92 & 0.29 & 0.07 & 9.56\\
    \\
Scenario 3 &  & 0.05 & 0.10 & {\bf 0.30} & 0.50 & 0.65 & 0.75\\
CRM & Selection (\%) & 0.01 & 11.01 & 72.39 & 16.41 & 0.18 & 0.00 & 16.59\\
    & Patient  (\%) & 3.68 & 6.71 & 17.20 & 7.45 & 0.91 & 0.05 & 8.41\\
New CRM & Selection (\%) & 0.01 & 11.96 & 73.77 & 14.17 & 0.09 & 0.00 & 14.26\\
    & Patient  (\%) & 1.78 & 7.26 & 18.57 & 7.38 & 0.89 & 0.12 & 8.39\\
        \\
Scenario 4 &  & 0.07 & 0.12 & 0.17 & {\bf 0.30} & 0.45 & 0.60\\
CRM & Selection (\%) & 0.06 & 1.56 & 22.43 & 57.80 & 17.55 & 0.60 & 18.15\\
    & Patient  (\%) & 4.08 &  4.51 & 8.87 & 12.46 & 5.33 & 0.76 & 6.08\\
New CRM & Selection (\%) & 0.01 & 1.90 & 22.61 & 58.27 & 16.69 & 0.52 & 17.21\\
    & Patient  (\%) & 2.03 & 3.21 & 9.40 & 14.14 & 5.91 & 1.30 & 7.21\\
        \\
Scenario 5 &  & 0.04 & 0.08 & 0.12 & 0.15 & {\bf 0.30} & 0.50\\
CRM & Selection (\%) & 0.00 & 0.13 & 1.84 & 21.62 & 63.01 & 13.40 & 13.40\\
    & Patient  (\%)  & 3.47 & 3.38 & 4.69 & 8.05 & 12.03 & 4.38 & 4.38\\
New CRM & Selection (\%) & 0.00 & 0.08 & 1.58 & 21.05 & 64.12 & 13.17 & 13.17\\
        & Patient  (\%) & 1.45 & 1.72 & 4.02 & 8.96 & 13.92 & 5.94 & 5.94\\
        \\
Scenario 6 &  & 0.05 & 0.14 & 0.18 & 0.20 & 0.23 & {\bf 0.30}\\
CRM & Selection (\%) & 0.01 & 2.06 & 10.32 & 22.33 & 34.26 & 31.02 & 0.00\\
    & Patient  (\%)  & 3.82 & 4.48 & 6.97 & 7.65 & 7.45 & 5.62 & 0.00\\
New CRM & Selection (\%) & 0.00 & 1.92 & 9.26 & 18.09 & 30.30 & 40.43 & 0.00\\
        & Patient  (\%) & 1.90 & 3.27 & 6.08 & 7.52 & 7.80 & 9.43 & 0.00\\
\hline
\end{tabular}
\end{table}

\begin{table}
\centering
\caption{Performance of CRM with unequal and fixed cohort sizes when total size is $N=42$. }
\label{table6}
\begin{tabular}{llrrrrrrr}
\hline
    &  & \multicolumn{6}{c}{Dose Level} &  \multicolumn{1}{c}{Overdose}\\
    & & & & & & & &    \multicolumn{1}{c}{Rate} \\
      \midrule
    &  & \multicolumn{1}{c}{1} & \multicolumn{1}{c}{2}
    & \multicolumn{1}{c}{3} & \multicolumn{1}{c}{4}
    & \multicolumn{1}{c}{5} & \multicolumn{1}{c}{6}\\
Scenario 1 &  &  {\bf 0.30} & 0.38 & 0.48 & 0.58 & 0.69 & 0.78\\
CRM & Selection (\%) & 69.98 & 26.75 & 3.18 & 0.10 & 0.00 & 0.00 & 30.02\\
    & Patient  (\%) &  28.02 & 10.24 & 3.18 & 0.51 & 0.04 & 0.00 & 13.98\\
New CRM & Selection (\%) &  68.99 & 27.47 & 3.45 & 0.09 & 0.00 & 0.00 & 31.01\\
    & Patient  (\%) & 25.51 & 10.81 & 4.37 & 1.15 & 0.14 & 0.02 & 16.49\\
    \\
Scenario 2 &  & 0.20 & {\bf 0.30} & 0.45 & 0.55 & 0.60 & 0.70\\
CRM & Selection (\%) & 24.09 & 60.58 & 14.81 & 0.51 & 0.02 & 0.00 & 15.34\\
    & Patient  (\%) & 15.22 & 17.75 & 7.59 & 1.28 & 0.15 & 0.01 & 9.03\\
New CRM & Selection (\%) & 24.50 & 60.20 & 14.76 & 0.53 & 0.01 & 0.00 & 15.30\\
    & Patient  (\%) & 13.66 & 17.66 & 8.31 & 2.00 & 0.31 & 0.06 & 10.68\\
    \\
Scenario 3 &  & 0.05 & 0.10 & {\bf 0.30} & 0.50 & 0.65 & 0.75\\
CRM & Selection (\%) & 0.00 & 9.54 & 76.39 & 13.97 & 0.10 & 0.00 & 14.07\\
    & Patient  (\%) & 3.71 & 7.36 & 21.57 & 8.34 & 0.97 & 0.05 & 9.37\\
New CRM & Selection (\%) & 0.01 & 10.87 & 76.19 & 12.89 & 0.05 & 0.00 & 12.94\\
    & Patient  (\%) & 1.84 & 8.08 & 22.83 & 8.21 & 0.92 & 0.13 & 9.26\\
        \\
Scenario 4 &  & 0.07 & 0.12 & 0.17 & {\bf 0.30} & 0.45 & 0.60\\
CRM & Selection (\%) & 0.01 & 1.12 & 20.52 & 62.05 & 15.96 & 0.33 & 16.29\\
    & Patient  (\%) & 4.07 & 4.55 & 10.19 & 16.02 & 6.38 & 0.80 & 7.18\\
New CRM & Selection (\%) & 0.01 & 1.01 & 20.69 & 62.73 & 15.34 & 0.22 & 15.56\\
    & Patient  (\%) & 2.02 & 3.30 & 10.57 & 17.75 & 7.06 & 1.31 & 8.36\\
        \\
Scenario 5 &  & 0.04 & 0.08 & 0.12 & 0.15 & {\bf 0.30} & 0.50\\
CRM & Selection (\%) & 0.00 & 0.04 & 1.27 & 19.83 & 67.22 & 11.63 & 11.63\\
    & Patient  (\%)  & 3.48 & 3.40 & 4.81 & 9.32 & 15.82  & 5.18 & 5.18\\
New CRM & Selection (\%) & 0.00 & 0.03 & 1.21 & 19.06 & 69.09 & 10.61 & 10.61\\
        & Patient  (\%) & 1.48 & 1.79 & 4.18 & 10.20 & 17.74 & 6.61 & 6.61\\
        \\
Scenario 6 &  & 0.05 & 0.14 & 0.18 & 0.20 & 0.23 & {\bf 0.30}\\
CRM & Selection (\%) & 0.00 & 1.23 & 9.31 & 21.56 & 33.46 & 34.44 & 0.00\\
    & Patient  (\%)  & 3.80 & 4.58 & 7.66 &  8.97 & 9.26  & 7.72 & 0.00\\
New CRM & Selection (\%) & 0.01 & 1.12 & 8.29 & 17.70 & 31.66 & 41.24 & 0.00\\
        & Patient  (\%) & 1.91 & 3.32 & 6.74 &  8.64 & 9.67 & 11.72 & 0.00\\
\hline
\end{tabular}
\end{table}

\begin{table}
\centering
\caption{Performance of Keyboard with unequal and fixed cohort sizes when total size is $N=24$. }
\label{table7}
\begin{tabular}{llrrrrrrr}
\hline
    &  & \multicolumn{6}{c}{Dose Level} &  \multicolumn{1}{c}{Overdose}\\
    & & & & & & & &    \multicolumn{1}{c}{Rate} \\
      \midrule
    &  & \multicolumn{1}{c}{1} & \multicolumn{1}{c}{2}
    & \multicolumn{1}{c}{3} & \multicolumn{1}{c}{4}
    & \multicolumn{1}{c}{5} & \multicolumn{1}{c}{6}\\
Scenario 1 &  &  {\bf 0.30} & 0.38 & 0.48 & 0.58 & 0.69 & 0.78\\
Keyboard  & Selection (\%) & 62.65 & 29.39 & 6.95 & 0.94 & 0.06 & 0.00 & 37.35\\
    & Patient  (\%) & 15.53 & 6.31 & 1.87 & 0.27 & 0.02 & 0.00 & 8.47\\
New Keyboard  & Selection (\%) & 59.97 & 29.00 & 9.80 & 1.18 & 0.05 & 0.00 & 40.03\\
    & Patient  (\%) & 12.64 & 6.98 & 3.45 & 0.78 & 0.14 & 0.00 & 11.36\\
    \\
Scenario 2 &  & 0.20 & {\bf 0.30} & 0.45 & 0.55 & 0.60 & 0.70\\
Keyboard  & Selection (\%) & 26.39 & 52.47 & 18.15 & 2.68 & 0.30 & 0.02 & 21.15\\
    & Patient  (\%) & 10.10 & 9.07 & 4.05 & 0.72 & 0.06 & 0.00 & 4.83\\
New Keyboard & Selection (\%)  & 27.43 & 48.21 & 21.41 & 2.67 & 0.27 & 0.00 & 24.36\\
    & Patient  (\%) & 7.78 & 8.99 & 5.64 & 1.31 & 0.27 & 0.01 & 7.23\\
    \\
Scenario 3 &  & 0.05 & 0.10 & {\bf 0.30} & 0.50 & 0.65 & 0.75\\
Keyboard  & Selection (\%) & 0.06 & 14.79 & 65.33 & 18.39 & 1.39 & 0.03 & 19.81\\
    & Patient  (\%) & 3.71 & 6.47 & 9.41 & 3.96 & 0.44 & 0.01 & 4.41\\
New Keyboard  & Selection (\%) & 0.15 & 11.83 & 71.73 & 15.55 & 0.74 & 0.01 & 16.30\\
    & Patient  (\%) & 1.70 & 5.79 & 10.74 & 4.81 & 0.93 & 0.03 & 5.77\\
        \\
Scenario 4 &  & 0.07 & 0.12 & 0.17 & {\bf 0.30} & 0.45 & 0.60\\
Keyboard & Selection (\%) & 0.25 & 6.43 & 26.96 & 45.89 & 18.28 & 2.18 & 20.47\\
    & Patient  (\%) & 4.10 & 5.23 & 6.32 & 5.78 & 2.25 & 0.31 & 2.56\\
New Keyboard  & Selection (\%) & 0.34 & 4.95 & 28.67 & 48.03 & 17.09 & 0.91 & 18.01\\
    & Patient  (\%) & 1.99 & 3.71 & 7.07 & 6.83 & 3.91 & 0.50 & 4.40\\
        \\
Scenario 5 &  & 0.04 & 0.08 & 0.12 & 0.15 & {\bf 0.30} & 0.50\\
Keyboard  & Selection (\%) & 0.01 & 1.54 & 7.56 & 28.36 & 47.36 & 15.17 & 15.17\\
    & Patient  (\%)  & 3.52 & 4.23 & 4.84 & 5.22 & 4.61 & 1.58 & 1.58\\
New Keyboard & Selection (\%) & 0.03 & 1.12 & 6.84 & 27.28 & 54.21 & 10.52 & 10.52\\
        & Patient  (\%) & 1.48 & 2.51 & 4.56 & 5.62 & 7.35 & 2.48 & 2.48\\
        \\
Scenario 6 &  & 0.05 & 0.14 & 0.18 & 0.20 & 0.23 & {\bf 0.30}\\
Keyboard & Selection (\%) & 0.18 & 9.15 & 17.73 & 23.48 & 26.45 & 23.02 & 0.00\\
    & Patient  (\%)  & 3.93 & 5.74 & 5.65 & 4.37 & 2.80 & 1.50  & 0.00\\
New Keyboard  & Selection (\%) & 0.25 & 6.68 & 17.14 & 22.78 & 27.90 & 25.25 & 0.00\\
        & Patient  (\%) & 1.95 & 4.01 & 5.66 & 4.61 & 4.55 & 3.21  & 0.00\\
\hline
\end{tabular}
\end{table}

\begin{table}
\centering
\caption{Performance of Keyboard with unequal and fixed cohort sizes when total size is $N=30$. }
\label{table8}
\begin{tabular}{llrrrrrrr}
\hline
    &  & \multicolumn{6}{c}{Dose Level} &  \multicolumn{1}{c}{Overdose}\\
    & & & & & & & &    \multicolumn{1}{c}{Rate} \\
      \midrule
    &  & \multicolumn{1}{c}{1} & \multicolumn{1}{c}{2}
    & \multicolumn{1}{c}{3} & \multicolumn{1}{c}{4}
    & \multicolumn{1}{c}{5} & \multicolumn{1}{c}{6}\\
Scenario 1 &  &  {\bf 0.30} & 0.38 & 0.48 & 0.58 & 0.69 & 0.78\\
Keyboard  & Selection (\%) & 66.48 & 26.37 & 6.46 & 0.67 & 0.02 & 0.00 & 33.52\\
    & Patient  (\%) &  19.23 & 7.96 & 2.42 & 0.37 & 0.03 & 0.00 & 10.77\\
New Keyboard  & Selection (\%) & 60.88 & 30.92 & 7.40 & 0.78 & 0.02 & 0.00 & 39.12\\
    & Patient  (\%) & 16.05 & 8.90 & 4.00 & 0.90 & 0.15 & 0.00 & 13.95\\
    \\
Scenario 2 &  & 0.20 & {\bf 0.30} & 0.45 & 0.55 & 0.60 & 0.70\\
Keyboard  & Selection (\%) & 27.21 & 52.06 & 18.22 & 2.29 & 0.20 & 0.02 & 20.72\\
    & Patient  (\%) & 11.72 & 11.76 & 5.46 & 0.97 & 0.09 & 0.01 & 6.52\\
New Keyboard & Selection (\%)  & 25.03 & 54.56 & 18.30 & 1.96 & 0.15 & 0.00 & 20.42\\
    & Patient  (\%) & 9.31 & 11.98 & 6.83 & 1.56 & 0.29 & 0.02 & 8.70\\
    \\
Scenario 3 &  & 0.05 & 0.10 & {\bf 0.30} & 0.50 & 0.65 & 0.75\\
Keyboard  & Selection (\%) & 0.02 & 11.00 & 71.17 & 16.92 & 0.86 & 0.02 & 17.80\\
    & Patient  (\%) & 3.72 & 7.49 & 12.76 & 5.39 & 0.62 & 0.02 & 6.03\\
New Keyboard  & Selection (\%) & 0.03 & 10.26 & 74.27 & 15.08 & 0.37 & 0.00 & 15.45\\
    & Patient  (\%) & 1.72 & 7.22 & 13.56 & 6.42 & 1.05 & 0.03 & 7.50\\
        \\
Scenario 4 &  & 0.07 & 0.12 & 0.17 & {\bf 0.30} & 0.45 & 0.60\\
Keyboard & Selection (\%) & 0.12 & 3.64 & 23.77 & 51.35 & 18.97 & 2.15 & 21.12\\
    & Patient  (\%) & 4.13 & 5.50 & 7.73 & 8.33 & 3.67 & 0.63 & 4.30\\
New Keyboard  & Selection (\%) & 0.09 & 3.68 & 22.93 & 54.60 & 17.91 & 0.78 & 18.69\\
    & Patient  (\%) & 2.02 & 4.02 & 8.41 & 9.64 & 5.21 & 0.70 & 5.91\\
        \\
Scenario 5 &  & 0.04 & 0.08 & 0.12 & 0.15 & {\bf 0.30} & 0.50\\
Keyboard  & Selection (\%) & 0.01 & 0.63 & 4.52 & 23.89 & 54.52 & 16.44 & 16.44\\
    & Patient  (\%)  & 3.51 & 4.29 & 5.19 & 6.57 & 7.27 & 3.17 & 3.17\\
New Keyboard & Selection (\%) & 0.00 & 0.65 & 4.07 & 22.90 & 60.32 & 12.05 & 12.05\\
        & Patient  (\%) & 1.48 & 2.59 & 4.85 & 7.21 & 10.12 & 3.75 & 3.75\\
        \\
Scenario 6 &  & 0.05 & 0.14 & 0.18 & 0.20 & 0.23 & {\bf 0.30}\\
Keyboard & Selection (\%) &  0.08 & 5.67 & 14.52 & 20.94 & 25.52 & 33.27 & 0.00\\
    & Patient  (\%)  & 3.98 & 6.12 & 6.55 & 5.57 & 4.24 & 3.54  & 0.00\\
New Keyboard  & Selection (\%) & 0.07 & 5.35 & 13.69 & 20.16 & 28.80 & 31.92 & 0.00\\
        & Patient  (\%) & 1.99 & 4.44 & 6.48 & 5.82 & 6.16 & 5.10  & 0.00\\
\hline
\end{tabular}
\end{table}

\begin{table}
\centering
\caption{Performance of Keyboard with unequal and fixed cohort sizes when total size is $N=36$. }
\label{table9}
\begin{tabular}{llrrrrrrr}
\hline
    &  & \multicolumn{6}{c}{Dose Level} &  \multicolumn{1}{c}{Overdose}\\
    & & & & & & & &    \multicolumn{1}{c}{Rate} \\
      \midrule
    &  & \multicolumn{1}{c}{1} & \multicolumn{1}{c}{2}
    & \multicolumn{1}{c}{3} & \multicolumn{1}{c}{4}
    & \multicolumn{1}{c}{5} & \multicolumn{1}{c}{6}\\
Scenario 1 &  &  {\bf 0.30} & 0.38 & 0.48 & 0.58 & 0.69 & 0.78\\
Keyboard  & Selection (\%) & 66.10 & 27.80 & 5.58 & 0.50 & 0.01 & 0.00 & 33.90\\
    & Patient  (\%) &  22.86 & 9.74 & 2.92 & 0.44 & 0.03 & 0.00 & 13.14\\
New Keyboard  & Selection (\%) & 62.74 & 30.22 & 6.62 & 0.41 & 0.01 & 0.00 & 37.26\\
    & Patient  (\%) & 19.73 & 10.64 & 4.57 & 0.92 & 0.14 & 0.00 & 16.27\\
    \\
Scenario 2 &  & 0.20 & {\bf 0.30} & 0.45 & 0.55 & 0.60 & 0.70\\
Keyboard  & Selection (\%) & 23.79 & 56.92 & 17.39 & 1.74 & 0.15 & 0.01 & 19.28\\
    & Patient  (\%) & 13.16 & 14.67 & 6.88 & 1.17 & 0.11 & 0.01 & 8.17\\
New Keyboard & Selection (\%)  & 23.20 & 57.02 & 18.33 & 1.34 & 0.11 & 0.00 & 19.78\\
    & Patient  (\%) & 11.00 & 14.62 & 8.36 & 1.69 & 0.30 & 0.02 & 10.38\\
    \\
Scenario 3 &  & 0.05 & 0.10 & {\bf 0.30} & 0.50 & 0.65 & 0.75\\
Keyboard  & Selection (\%) & 0.00 & 8.49 & 77.11 & 13.88 & 0.50 & 0.01 & 14.40\\
    & Patient  (\%) & 3.74 & 8.46 & 16.15 & 6.87 & 0.76 & 0.03 & 7.65\\
New Keyboard  & Selection (\%) & 0.01 & 7.77 & 79.59 & 12.33 & 0.30 & 0.00 & 12.63\\
    & Patient  (\%) & 1.73 & 8.05 & 17.58 & 7.44 & 1.17 & 0.03 & 8.65\\
        \\
Scenario 4 &  & 0.07 & 0.12 & 0.17 & {\bf 0.30} & 0.45 & 0.60\\
Keyboard & Selection (\%) & 0.04 & 2.18 & 22.23 & 54.40 & 19.63 & 1.52 & 21.15\\
    & Patient  (\%) & 4.14 & 5.66 & 9.06 & 11.13 & 5.12 & 0.89 & 6.01\\
New Keyboard  & Selection (\%) & 0.05 & 2.43 & 22.66 & 56.65 & 17.41 & 0.81 & 18.22\\
    & Patient  (\%) & 2.03 & 4.20 & 10.19 & 12.15 & 6.55 & 0.88 & 7.43\\
        \\
Scenario 5 &  & 0.04 & 0.08 & 0.12 & 0.15 & {\bf 0.30} & 0.50\\
Keyboard  & Selection (\%) & 0.00 & 0.21 & 2.70 & 20.29 & 61.75 & 15.05 & 15.05\\
    & Patient  (\%)  & 3.52 & 4.31 & 5.37 & 7.81 & 10.32 & 4.67 & 4.67\\
New Keyboard & Selection (\%) & 0.00 & 0.36 & 2.93 & 19.63 & 65.78 & 11.31 & 11.31\\
        & Patient  (\%) & 1.49 & 2.62 & 5.10 & 8.57 & 13.27 & 4.95 & 4.95\\
        \\
Scenario 6 &  & 0.05 & 0.14 & 0.18 & 0.20 & 0.23 & {\bf 0.30}\\
Keyboard & Selection (\%) & 0.02 & 3.78 & 11.96 & 18.38 & 26.33 & 39.53 & 0.00\\
    & Patient  (\%)  & 3.98 & 6.39 &  7.31 & 6.68 & 5.70 & 5.94  & 0.00\\
New Keyboard  & Selection (\%) & 0.03 & 4.10 & 11.97 & 18.13 & 28.08 & 37.69 & 0.00\\
        & Patient  (\%) & 2.03 & 4.72 & 7.24 & 6.91 & 7.77 & 7.34 & 0.00\\
\hline
\end{tabular}
\end{table}

\begin{table}
\centering
\caption{Performance of Keyboard with unequal  and fixed cohort sizes when total size is $N=42$.}
\label{table10}
\begin{tabular}{llrrrrrrr}
\hline
    &  & \multicolumn{6}{c}{Dose Level} &  \multicolumn{1}{c}{Overdose}\\
    & & & & & & & &    \multicolumn{1}{c}{Rate} \\
      \midrule
    &  & \multicolumn{1}{c}{1} & \multicolumn{1}{c}{2}
    & \multicolumn{1}{c}{3} & \multicolumn{1}{c}{4}
    & \multicolumn{1}{c}{5} & \multicolumn{1}{c}{6}\\
Scenario 1 &  &  {\bf 0.30} & 0.38 & 0.48 & 0.58 & 0.69 & 0.78\\
Keyboard & Selection (\%) & 67.81 & 27.05 & 4.81 &  0.31 & 0.02 & 0.00 & 32.19\\
    & Patient  (\%) & 26.72 & 11.32 & 3.40 & 0.53 & 0.04  &  0.00 & 15.28\\
New Keyboard & Selection (\%) & 64.54 & 29.93 & 5.22 & 0.31 & 0.00 & 0.00 & 35.46\\
    & Patient  (\%) & 23.63 & 12.35 & 4.88 & 0.98 & 0.16 & 0.00 & 18.37\\
    \\
Scenario 2 &  & 0.20 & {\bf 0.30} & 0.45 & 0.55 & 0.60 & 0.70\\
Keyboard & Selection (\%) & 21.96 & 59.45 & 17.09 & 1.37 & 0.13 & 0.00 & 18.59\\
    & Patient  (\%) & 14.42 & 17.66  & 8.35 & 1.41 & 0.15 & 0.01 & 9.92\\
New Keyboard & Selection (\%) & 22.25 & 59.52 & 17.31 & 0.83 & 0.09 & 0.00 & 18.23\\
    & Patient  (\%) & 12.34 & 17.80 & 9.64 & 1.85 & 0.35 & 0.02 & 11.87\\
    \\
Scenario 3 &  & 0.05 & 0.10 & {\bf 0.30} & 0.50 & 0.65 & 0.75\\
Keyboard & Selection (\%) & 0.00 & 6.80 & 80.69 & 12.18 & 0.31 & 0.02 & 12.51\\
    & Patient  (\%) & 3.74 & 9.48 & 19.77 & 8.13 & 0.84 & 0.04 & 9.01\\
New Keyboard & Selection (\%) & 0.00 & 5.96 & 83.12 & 10.80 & 0.12 & 0.00 & 10.92\\
    & Patient  (\%) & 1.69 & 9.30 & 20.76 & 8.95 & 1.27 & 0.03 & 10.25\\
        \\
Scenario 4 &  & 0.07 & 0.12 & 0.17 & {\bf 0.30} & 0.45 & 0.60\\
Keyboard & Selection (\%) & 0.02 & 1.63 & 19.29 & 59.04 & 18.88 & 1.14  & 20.02\\
    & Patient  (\%) & 4.20 & 5.83 & 10.28 & 14.03  & 6.53 & 1.14 & 7.66\\
New Keyboard & Selection (\%) & 0.02 & 1.30 & 21.53 & 61.26 & 15.35 & 0.54 & 15.89\\
    & Patient  (\%) & 2.04 & 4.37 & 11.66 & 15.30 & 7.61 & 1.01 & 8.62\\
        \\
Scenario 5 &  & 0.04 & 0.08 & 0.12 & 0.15 & {\bf 0.30} & 0.50\\
Keyboard & Selection (\%) & 0.00 & 0.13 & 1.47 & 17.74 & 67.97 & 12.69 & 12.69\\
    & Patient  (\%)  & 3.50 & 4.29 & 5.39 & 9.01 & 13.65 & 6.15 & 6.15\\
New Keyboard & Selection (\%) & 0.00 & 0.11 & 1.94 & 17.94 & 70.01 & 10.00 & 10.00\\
        & Patient  (\%) & 1.49 & 2.59 & 5.17 & 9.94 & 16.51 & 6.30 & 6.30\\
        \\
Scenario 6 &  & 0.05 & 0.14 & 0.18 & 0.20 & 0.23 & {\bf 0.30}\\
Keyboard & Selection (\%) & 0.01 & 2.54  & 9.50 & 16.63 & 27.51 & 43.81 & 0.00\\
    & Patient  (\%)  & 3.99 & 6.59 & 7.78 & 7.75 & 7.36 & 8.52 & 0.00\\
New Keyboard & Selection (\%) & 0.00 & 2.79 & 10.95 & 16.74 & 28.79 & 40.73 & 0.00\\
        & Patient  (\%) & 2.02 & 5.01 & 7.93 & 7.87 & 9.34 & 9.83 & 0.00\\
\hline
\end{tabular}
\end{table}

\spacingset{1.9} 
\section{Conclusion}
This manuscript discusses  the design of Phase I clinical trials with unequal cohort sizes.
The proposed cohort sizes can be easily incorporated into existing designs.
The new design may perform worse than using a fixed cohort size if the lowest dosage is the MTD.
Usually, it is not a problem because in most trials, the lowest dose is much less toxic than the target level, so it’s seldom the MTD.
For other cases,  using unequal cohort sizes is allays better than using fixed cohort sizes in terms of
saving trial duration and improving the accuracy of MTD recommendations.
From the Fisher information viewpoint, treating patients with a dose that has a $50\%$ chance of being toxic provides the most information.
But, it’s normally not ethically right.
Taking patients safety into account, treating them at the MTD can benefit patients and provide more information.
Note that the first two cohorts each have only one person in the new design. 
Beginning with smaller cohort sizes at low doses,  allows more patients to eventually receive higher and appropriate doses.
This provides more information and therefore improving recommendation accuracy in turn.
Despite the risk of overdosing, simulations show that losses are usually limited to 1-2 patients compared to the traditional fixed cohort sizes.

This manuscript studies the performance of using unequal cohort sizes in Phase I trials.
It will be of interest to further apply and evaluate the proposed unequal cohort sizes to a Phase II design or a Phase I/II design \cite{Yua5}.
We are currently working on this and hope to report our findings in a future paper.

\bibliographystyle{Chicago}
\bibliography{ref.bib}

\begin{thebibliography}{}

\bibitem[\protect\citeauthoryear{Babb, Rogatko, and Zacks}{Babb
  et~al.}{1998}]{Bab}
Babb, J., A.~Rogatko, and S.~Zacks (1998).
\newblock Cancer phase {I} clinical trials: efficient dose escalation with
  overdose control.
\newblock {\em Statistics in Medicine\/}~{\em 17}, 1103--1120.

\bibitem[\protect\citeauthoryear{Braun}{Braun}{2002}]{Bra}
Braun, T. (2002).
\newblock The bivariate continual reassessment method: extending the crm to
  phase {I} trials of two competing outcomes.
\newblock {\em Controlled Clinical Trials\/}~{\em 23}, 240--256.

\bibitem[\protect\citeauthoryear{Chen, Peace, and Zhang}{Chen
  et~al.}{2020}]{Che3}
Chen, D., K.~Peace, and P.~Zhang (2020).
\newblock {\em Clinical Trial Data Analysis Using R and SAS\/} (2nd ed.).
\newblock Boca Raton, FL: CRC Press.

\bibitem[\protect\citeauthoryear{Cheung}{Cheung}{2011}]{Che2}
Cheung, Y. (2011).
\newblock {\em Dose Finding by the Continual Reassessment Method}.
\newblock Boca Raton, FL: CRC Press.

\bibitem[\protect\citeauthoryear{Cheung and Chappel}{Cheung and
  Chappel}{2002}]{Che4}
Cheung, Y. and R.~Chappel (2002).
\newblock A simple technique to evaluate model sensitivity in the continual
  reassessment method.
\newblock {\em Biometrics\/}~{\em 58}, 671--674.

\bibitem[\protect\citeauthoryear{Cheung and Chappell}{Cheung and
  Chappell}{2000}]{Che}
Cheung, Y. and R.~Chappell (2000).
\newblock Sequential designs for phase {I} clinical trials with late-onset
  toxicities.
\newblock {\em Statistics in Medicine\/}~{\em 56}, 1177--1182.

\bibitem[\protect\citeauthoryear{Durham, Flournoy, and Rosenberger}{Durham
  et~al.}{1997}]{Dur}
Durham, S., N.~Flournoy, and W.~Rosenberger (1997).
\newblock A random walk rule for phase {I} clinical trials.
\newblock {\em Biometrics\/}~{\em 53}, 745--760.

\bibitem[\protect\citeauthoryear{Guo, Wang, Yang, Lynn, and Ji}{Guo
  et~al.}{2017}]{Guo}
Guo, W., S.~Wang, S.~Yang, H.~Lynn, and Y.~Ji (2017).
\newblock A bayesian interval dose-finding design addressing ockham's razor:
  mtpi-2.
\newblock {\em Contemporary Clinical Trials\/}~{\em 58}, 23--33.

\bibitem[\protect\citeauthoryear{Horton, Wages, and Conaway}{Horton
  et~al.}{2017}]{Hor}
Horton, B., N.~Wages, and M.~Conaway (2017).
\newblock Performance of toxicity probability interval based designs in
  contrast to the continual reassessment method.
\newblock {\em Statistics in Medicine\/}~{\em 36}, 291--300.

\bibitem[\protect\citeauthoryear{Ivanova, Montazer-Haghighi, Mohanty, and
  Durham}{Ivanova et~al.}{2003}]{Iva}
Ivanova, A., A.~Montazer-Haghighi, S.~Mohanty, and S.~Durham (2003).
\newblock Improved up-and-down designs for phase {I} trials.
\newblock {\em Statistics in Medicine\/}~{\em 22}, 69--82.

\bibitem[\protect\citeauthoryear{Ji, Liu, Li, and Bekele}{Ji et~al.}{2010}]{Ji}
Ji, Y., P.~Liu, Y.~Li, and B.~Bekele (2010).
\newblock A modified toxicity probability interval method for dose-finding
  trials.
\newblock {\em Clinical Trials\/}~{\em 7}, 235--244.

\bibitem[\protect\citeauthoryear{Kojima}{Kojima}{2021}]{Koj}
Kojima, M. (2021).
\newblock Early completion of phase {I} cancer clinical trials with bayesian
  optimal interval design.
\newblock {\em Statistics in Medicine\/}~{\em 40}, 3215--3226.

\bibitem[\protect\citeauthoryear{Lee, Wages, Goodman, and Lockhart}{Lee
  et~al.}{2021}]{Lee2021}
Lee, S., N.~Wages, K.~Goodman, and A.~Lockhart (2021).
\newblock Designing dose-finding phase {I} clinical trials: top 10 questions
  that should be discussed with your statistician.
\newblock {\em JCO precision oncology\/}~{\em 5}, 317--324.

\bibitem[\protect\citeauthoryear{Li and Yuan}{Li and Yuan}{2020}]{Li}
Li, Y. and Y.~Yuan (2020).
\newblock Pa-crm: A continuous reassessment method for pediatric phase {I}
  oncology trials with concurrent adult trials.
\newblock {\em Biometrics\/}~{\em 76}, 1364--1373.

\bibitem[\protect\citeauthoryear{Lin and Yin}{Lin and Yin}{2017}]{Lin}
Lin, R. and G.~Yin (2017).
\newblock Bayesian optimal interval design for dose finding in drug-combination
  trials.
\newblock {\em Statistical Methods in Medical Research\/}~{\em 26}, 2155--2167.

\bibitem[\protect\citeauthoryear{Lin and Yuan}{Lin and Yuan}{2020}]{Lin2}
Lin, R. and Y.~Yuan (2020).
\newblock Time-to-event model-assisted designs for dose-finding trials with
  delayed toxicity.
\newblock {\em Biostatistics\/}~{\em 21}, 807--824.

\bibitem[\protect\citeauthoryear{Liu, Wang, and Ji}{Liu et~al.}{2020}]{Liu3}
Liu, M., S.~Wang, and Y.~Ji (2020).
\newblock The i3+3 design for phase {I} clinical trials.
\newblock {\em Journal of Biopharmaceutical Statistics\/}~{\em 30}, 294--304.

\bibitem[\protect\citeauthoryear{Liu, Yin, and Yuan}{Liu et~al.}{2013}]{Liu}
Liu, S., G.~Yin, and Y.~Yuan (2013).
\newblock Bayesian data augmentation dose finding with continual reassessment
  method and delayed toxicity.
\newblock {\em Annals of Applied Statistics\/}~{\em 4}, 2138--2156.

\bibitem[\protect\citeauthoryear{Liu and Yuan}{Liu and Yuan}{2015}]{Liu2}
Liu, S. and Y.~Yuan (2015).
\newblock Bayesian optimal interval designs for phase {I} clinical trials.
\newblock {\em Journal of the Royal Statistical Society, Series C (Applied
  Statistics)\/}~{\em 64}, 507--523.

\bibitem[\protect\citeauthoryear{O'Connell, Wages, and Garrett-Mayer}{O'Connell
  et~al.}{2023}]{O'co}
O'Connell, N., N.~Wages, and E.~Garrett-Mayer (2023).
\newblock Quasi-partial order continual reassessment method: Applying toxicity
  scores to cancer dose-finding drug combination trials.
\newblock {\em Biometrics\/}~{\em 125}, 107050.

\bibitem[\protect\citeauthoryear{O'Quigley, Pepe, and Fisher}{O'Quigley
  et~al.}{1990}]{O'qu}
O'Quigley, J., M.~Pepe, and L.~Fisher (1990).
\newblock Continual reassessment method: a practical design for phase {I}
  clinical trials in cancer.
\newblock {\em Biometrics\/}~{\em 46}, 33--48.

\bibitem[\protect\citeauthoryear{Petroni, Wages, Paux, and Dubois}{Petroni
  et~al.}{2018}]{Pet}
Petroni, G., N.~Wages, G.~Paux, and F.~Dubois (2018).
\newblock Implementation of adaptive methods in early-phase clinical trials.
\newblock {\em Statistics in Medicine\/}~{\em 36}, 215--224.

\bibitem[\protect\citeauthoryear{Rogatko, Schoeneck, Jonas, Tighiouart, Khuri,
  and Porter}{Rogatko et~al.}{2007}]{Rog}
Rogatko, A., D.~Schoeneck, W.~Jonas, M.~Tighiouart, F.~Khuri, and A.~Porter
  (2007).
\newblock Translation of innovative designs into phase {I} trials.
\newblock {\em Journal of Clinical Oncology\/}~{\em 25}, 4982--4986.

\bibitem[\protect\citeauthoryear{Shen and O'Quigle}{Shen and
  O'Quigle}{1996}]{She}
Shen, L. and J.~O'Quigle (1996).
\newblock Consistency of continual reassessment method under model
  misspecification.
\newblock {\em Biometrika\/}~{\em 83}, 395--405.

\bibitem[\protect\citeauthoryear{Simon, Freidlin, Rubinstein, Arbuck, Collins,
  and Christian}{Simon et~al.}{1997}]{Sim}
Simon, R., B.~Freidlin, L.~Rubinstein, S.~Arbuck, J.~Collins, and M.~Christian
  (1997).
\newblock Accelerated titration designs for phase {I} clinical trials in
  oncology.
\newblock {\em Journal of the National Cancer Institute\/}~{\em 89},
  1138--1147.

\bibitem[\protect\citeauthoryear{Skolnik, Barrett, Jayaraman, Patel, and
  Adamson}{Skolnik et~al.}{2008}]{Sko}
Skolnik, J., J.~Barrett, B.~Jayaraman, D.~Patel, and P.~Adamson (2008).
\newblock Shortening the timeline of pediatric phase {I} trials: the rolling
  six design.
\newblock {\em Journal of Clinical Oncology\/}~{\em 26}, 190--195.

\bibitem[\protect\citeauthoryear{Storer}{Storer}{1989}]{Sto}
Storer, B. (1989).
\newblock Design and analysis of phase {I} clinical trials.
\newblock {\em Biometrics\/}~{\em 45}, 925--937.

\bibitem[\protect\citeauthoryear{Stylianou and Follmann}{Stylianou and
  Follmann}{2004}]{Sty}
Stylianou, M. and D.~Follmann (2004).
\newblock The accelerated biased coin up‐and‐down design in phase {I}
  trials.
\newblock {\em Journal of Biopharmaceutical Statistics\/}~{\em 14}, 249--260.

\bibitem[\protect\citeauthoryear{Wages, Conaway, and O'Quigley}{Wages
  et~al.}{2011}]{Wag}
Wages, N., M.~Conaway, and J.~O'Quigley (2011).
\newblock Continual reassessment method for partial ordering.
\newblock {\em Biometrics\/}~{\em 67}, 1555--1563.

\bibitem[\protect\citeauthoryear{Yan, Mandrekar, and Yuan}{Yan
  et~al.}{2017}]{Yan}
Yan, F., S.~Mandrekar, and Y.~Yuan (2017).
\newblock Keyboard: a novel bayesian toxicity probability interval design for
  phase {I} clinical trials.
\newblock {\em Clinical Cancer Research\/}~{\em 23}, 3994--4993.

\bibitem[\protect\citeauthoryear{Yin}{Yin}{2012}]{Yin2}
Yin, G. (2012).
\newblock {\em Clinical Trial Design: Bayesian and Frequentist Adaptive
  Methods}.
\newblock Hoboken, New Jersey: Wiley.

\bibitem[\protect\citeauthoryear{Yin and Yuan}{Yin and Yuan}{2009}]{Yin}
Yin, G. and Y.~Yuan (2009).
\newblock Bayesian model averaging continual reassessment method in phase {I}
  clinical trials.
\newblock {\em Journal of the American Statistical Association\/}~{\em 104},
  954--968.

\bibitem[\protect\citeauthoryear{Yuan, Hess, Hilsenbeck, and Gilbert}{Yuan
  et~al.}{2016}]{Yua}
Yuan, Y., K.~Hess, S.~Hilsenbeck, and M.~Gilbert (2016).
\newblock Bayesian optimal interval design: a simple and well-performing design
  for phase {I} oncology trials.
\newblock {\em Clinical Cancer Research\/}~{\em 22}, 4291--4301.

\bibitem[\protect\citeauthoryear{Yuan, Lin, Li, Nie, and Warren}{Yuan
  et~al.}{2018}]{Yua2}
Yuan, Y., R.~Lin, D.~Li, L.~Nie, and K.~Warren (2018).
\newblock Time-to-event bayesian optimal interval design to accelerate phase
  {I} trials.
\newblock {\em Clinical Cancer Research\/}~{\em 24}, 4921--4930.

\bibitem[\protect\citeauthoryear{Yuan, Nguyen, and Thall}{Yuan
  et~al.}{2021}]{Yua5}
Yuan, Y., H.~Nguyen, and P.~Thall (2021).
\newblock {\em Bayesian Designs for Phase {I}-{II} Clinical Trials}.
\newblock Boca Raton, FL: CRC Press.

\bibitem[\protect\citeauthoryear{Zhou, Murray, Pan, and Yuan}{Zhou
  et~al.}{2018}]{Zho}
Zhou, H., T.~Murray, H.~Pan, and Y.~Yuan (2018).
\newblock Comparative review of novel model-assisted designs for phase {I}
  clinical trials.
\newblock {\em Statistics in Medicine\/}~{\em 37}, 2208--2222.

\end{thebibliography}
\end{document}